# Mind the gap: Bayesian equipoise calibration of clinical trial designs

**Fabio Rigat, PhD***

**Abstract**

A key objective of randomized clinical trial design is to ensure strong control of the conditional error rates associated with the primary analysis outcome, but no link between trial design and the probabilities of the design hypotheses is currently established. This paper contributes to bridging this gap by calibrating the operational characteristics of the trial design with respect to the pre-specified level of equipoise imbalance provided by the primary trial outcome, which is a reduction in pre-study uncertainty about whether the null or the alternative hypothesis are more likely true. First, equipoise imbalance is quantified as the percentile of the post-study odds of the design hypotheses on a prior distribution reflecting pre-study equipoise within the population of medical experts. Under this prior, common late phase designs are shown to provide at least 90% evidence of equipoise imbalance. Designs carrying 95% power at 5% false positive rate are also shown to provide strong equipoise imbalance against the alternative hypothesis when the trial outcome fails to reject the null, providing a robust statistical basis to informing further clinical development decisions. Finally, equipoise calibration is applied to design of sequential clinical development plans using oncology efficacy endpoints. Commonly used power and false positive error rates are shown to provide strong equipoise imbalance when positive outcomes are observed in both phase 2 and phase 3. Establishing strong equipoise imbalance based on inconsistent outcomes of phase 2 and phase 3 studies is shown to require large sample sizes unlikely to improve current evidence standards.

* AstraZeneca Oncology Biometrics, Cambridge, UK; fabio.rigat@astrazeneca.com

## Introduction

The design and analysis of clinical trials are guided by harmonised principles [1], with further recent developments in several areas including paediatric extrapolation, design adaptation and consideration of patient preferences [2-4]. The application of these principles ensures that the precision associated with primary trial outcomes conveys appropriate statistical evidence. However, precision per se is insufficient for a statistically positive trial outcome to be practice-changing, because p-values and posterior probabilities do not necessarily map by design into clinically meaningful treatment effects [5]. The estimands addendum to the ICH E9 guideline on statistical principles for clinical trials [6] in fact recognized that the interpretation and practical impact of clinical trial outcomes hinges on multiple factors, including the magnitude and precision of the observed treatment effects.

From a clinical perspective, a clinical trial outcome can be practice-changing if it demonstrates a *clinical equipoise imbalance*, that is a reduction in a pre-study "state of genuine uncertainty within the expert medical community about the preferred treatment" [7-8]. The practical relevance of equipoise was recently highlighted in oncology [9-10]. Here a perceived equipoise imbalance emerging during conduct of an open label randomized controlled study was recognized as a driver of asymmetric dropouts and to atypical practices when monitoring discrepancies between clinical assessments by investigators and by independent blinded reviewers. Hence, lack of equipoise considerations at design stage represents a gap between sample size calculations and a key attribute for trial outcomes to be recognized as practice changing. The approach presented here contributes to reducing this gap by providing a definition of *equipoise calibration* of the trial design, under which the statistical properties of the primary study outcome are interpretable from a clinical equipoise perspective. Specifically, a formal Bayesian definition of clinical equipoise is presented, linking pre-specified frequentist properties of the trial outcome to a population model of equipoise, thus mitigating potential gaps between statistically and clinically significant trial outcomes. To this end, clinical equipoise is recognized as a direct statement about the probabilities of competing clinical hypotheses among relevant experts. A positive outcome of an equipoise calibrated study is not only highly unlikely under the null hypothesis, but it also needs to be associated with post-study odds of the design hypotheses strongly inconsistent with their pre-study equipoise probability distribution.

The paper is organized as follows: Sections 1 and 2 discuss three statistical models of clinical equipoise for study design. The equipoise model based on the least pre-study information about the trial hypotheses is motivated as a reference for practical design work. Under this model, common late phase designs are shown to provide evidence of equipoise imbalance at least at the 90$^{th}$ percentile of the pre-study equipoise distribution. Superiority designs carrying 95% power at 5% false positive rate are also shown to provide strong equipoise imbalance against the alternative hypothesis when the trial outcome fails to reject the null, which provides a robust statistical basis to informing further clinical development decisions. Section 3 applies equipoise calibration to design of sequential clinical development plans comprising one randomised phase 2 trial followed by one phase 3 confirmatory trial, using efficacy endpoints and data analysis models commonly used in oncology. Typical designs are shown to provide high overall levels of equipoise imbalance in favour of the joint phase 2 and phase 3 design hypotheses when positive trial outcomes are observed in both studies. Large increases in sample sizes are shown to be insufficient to providing strong equipoise imbalance in favour of the joint null hypothesis when a phase 3 trial fails to confirm a positive randomised phase 2 outcome. The discussion examines the application of equipoise calibration in practice, including instances when perfect pre-study clinical equipoise is unlikely to hold.

## 1. The post-study odds of the design hypotheses quantify clinical equipoise imbalance

Let the relative likelihood of the null hypothesis $H_0$ versus the motivating trial hypothesis $H_1$ be quantified by their pre-study odds, $\frac{P(H_0)}{P(H_1)}$ [11]. Using the odds form of Bayes theorem, the post-study odds of the design hypotheses are the product between pre-study odds and the likelihood ratios of the study outcome,

$$r_{10}(+) \coloneqq \frac{P(H_1 \mid +)}{P(H_0 \mid +)} = \frac{P(H_1)}{P(H_0)} \times \frac{p(+\mid H_1)}{p(+\mid H_0)}, \qquad (1)$$

$$r_{01}(-) \coloneqq \frac{P(H_0 \mid -)}{P(H_1 \mid -)} = \frac{P(H_0)}{P(H_1)} \times \frac{p(-\mid H_0)}{p(-\mid H_1)}. \qquad (2)$$

The symbols “+” and “–“ in (1)-(2) respectively represent statistically positive and negative analysis outcomes, as determined by a pre-specified effect size [12]. The examples provided in this paper focus in on randomized designs appropriate to establishing superiority in a relevant clinical endpoint measured in participants assigned to one investigational arm or to a parallel standard of care arm. Negative outcomes here are intended as failing to achieve statistical superiority, as opposed to showing inferior clinical outcomes to standard of care.

The operational characteristics $p(+\mid H_0)$ and $p(+\mid H_1)$ are respectively the false positive error rate and power of the analysis outcomes, and $r_{10}(+)$ and $r_{01}(-)$ in (1)-(2) represent the post-study odds in favour of $H_1$ or of $H_0$ should the outcome be respectively positive or negative. Specific examples of (1)-(2) are provided here using established clinical response criteria, endpoints and statistical models used in oncology clinical development [13-19].

The post-study odds in (1)-(2) combine two sources of uncertainty, represented by lowercase "$p$" and by capital "$P$" respectively. The uncertainty $p$ represents the frequentist operational characteristics of the statistical analysis outcome, including power and false positive error rate, which also depend on study sample size. These are the objective, conditionally self-standing components of the post-study odds routinely used for sample size calculation in practice. Ratios of capital $P$s in (1)-(2) represent epistemic probability [20] under which the operating characteristics $p$ are taken to be valid, including the definition of meaningful primary endpoints and the expected magnitude of the treatment effects defining the trial hypothesis $H_1$.

By (1)-(2), a positive study outcome carrying 90% power at 5% false positive error rate increases the pre-trial odds of $H_1$ versus $H_0$ by a factor of $0.9/0.05 = 18$. Likewise, a negative study outcome increases the pre-trial odds of $H_0$ versus $H_1$ by a factor of $(1 - 0.05)/0.1 = 9.5$. Determining the post-trial odds levels implied by these changes requires a description of the distribution of the pre-study odds, which can be quantified using formal elicitation methods [21-26]. Elicitation methods rely on probabilistic judgements collected from a sample of subject matter experts to approximate the population distribution of relevant features of the study design, including the primary endpoint effect size. Rather than relying on elicitation, this work explores the impact on trial sample size when a positive trial outcome is required to be both statistically significant and to demonstrate strong clinical equipoise imbalance, as measured by the percentile of the post-study odds (1) on a pre-study odds population distribution under clinical equipoise. To define a pre-study odds equipoise distribution, it is useful to consider that odds of 2:1 do not reflect robust equipoise imbalance according to the only survey quantifying pre-study uncertainty within an expert medical community published to date [27]. Also, statistical modelling shows that 3:1 odds are associated to barely significant p-values at the traditional 5% false positive rate level, while odds greater than 14:1 correspond to p-values smaller than 0.005 and are recommended as a more robust basis for discovery claims [28]. Building on these results, the next section explores three probabilistic models of clinical equipoise.

## 2. Three probabilistic models of clinical equipoise for calibration of clinical trial design

The first probability model of the pre-study odds population distribution considered here is

$$P(H_1) \coloneqq 1 - P(H_0), \qquad (3)$$

$$P(H_0) \sim U(0,1). \qquad (4)$$

Equation (3) states that the pre-study evidence in favour of the design hypothesis is defined as the complement to any available evidence in favour of the null. Among all the epistemic probability distributions which may be used to describe pre-study subjective confidence $P(H_0)$ about the truth of $H_0$ among medical experts, the uniform distribution (4) assigns equal prior weight to any possible value $P(H_0) = p, p \in (0,1)$.

Prior (3) has been amply discussed in the literature, its main strengths including its probabilistic representation of the principle of insufficient reason and its entropy maximization property [29]. The main limitations of the uniform prior have also been amply discussed, especially in relation to its lack of invariance [30-32], implying that the odds $P(H_0)/(1 - P(H_0))$ are not uniformly distributed when $P(H_0)$ is uniform. Under (3)-(4) the analytical form of the pre-study odds probability distribution is in fact Beta Prime with parameters (1,1), hereby referred to as $BP(1,1)$.

Availability of the analytical form of the pre-study odds distribution is a valuable property for studying its applicability to quantifying equipoise imbalance. Figure 1 depicts the cumulative distribution function of the $BP(1,1)$ model, alongside to those of two alternative models. Each curve in Figure 1 depicts the cumulative proportion of the expert medical population associated by each clinical equipoise model with pre-study odds smaller or equal to the values shown on the horizontal axis. The $BP(1,1)$ and $BP(0.5,0.5)$ models, depicted respectively in black and red in Figure 1, represent clinical equipoise as median pre-study odds equal to 1:1. The $BP(0.5,0.5)$ model describes the population distribution of $P(H_0)$ as $Beta(0.5,0.5)$ and equally concentrated towards its extreme values 0 or 1. Model $BP(1,2)$, depicted in blue in Figure 1, replaces the uniform distribution (4) with a $Beta(1,2)$, which represents clinical equipoise as population average pre-study odds of 1:1.

*2.2 Application of equipoise models to study design calibration*

By (1), a positive outcome carrying 90% power at 5% false positive error rate, is associated with post-study odds in favour of $H_1$ equal to 0.9/0.05 = 18:1 for an individual in perfect pre-study equipoise. Under the $BP(1,1)$ model (3)-(4), the percentile of the 18:1 odds within the population in pre-study equipoise is 18/(1+18) ≈ 94.7%. Hence, under (3)-(4) observing a positive outcome at 90% power at 5% false positive rate will demonstrate clinical equipoise imbalance at the 94.7$^{th}$ percentile of the pre-study equipoise distribution to any individual in perfect pre-study equipoise. More generally, establishing clinical equipoise imbalance based on the 95$^{th}$ percentiles of the *BP(1,2), BP(1,1)* and *BP(0.5,0.5)* pre-study odds distributions in Figure 1 requires post-study odds greater than 3.5:1, 19:1 and 161:1 respectively.

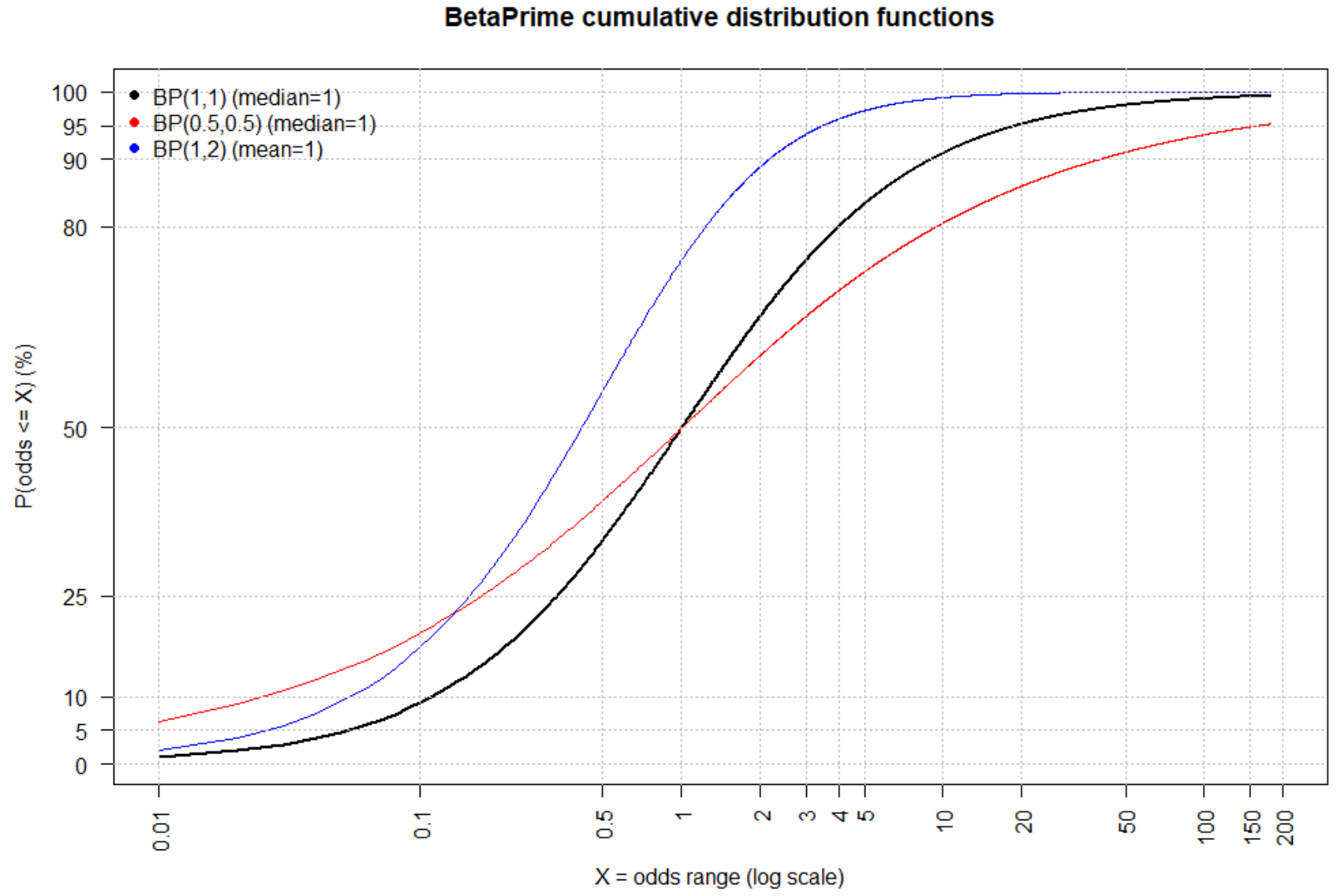

*Figure 1: cumulative distribution functions of three pre-study odds equipoise models. The BP(1,1) model, depicted in black, assumes that pre-study evidence in favour of the null hypothesis is uniformly distributed within the expert clinical population. The BP(0.5,0.5) model shown in red represents pre-study evidence concentrated towards* $P(H_0) = 0$ *or* $P(H_0) = 1$*. Model BP(1,2) shown in blue assumes weak evidence against the null distribution. Clinical equipoise is represented as median pre-study odds of 1 by the BP(1,1) and BP(0.5,0.5) models, and by average pre-study odds of 1 by the BP(1,2) model.*

These numerical discrepancies show that none of the models depicted in Figure 1 should be taken as prescriptive, because clinical equipoise admits multiple probabilistic representations associated with different post-study odds thresholds representing strong equipoise imbalance.

Model *BP(1,1)* is proposed here as a reference equipoise representation for trial design calibration based on three rationales. First, the *BP(1,1)* distribution assumes minimal precision of pre-study evidence in favour or against the null hypothesis, which maximises its range of potential applications in early and late clinical development. Second, the post-study odds associated to equipoise imbalance under the *BP(1,1)* model are no weaker than current practice for confirmatory trial design, unlike for the *BP(1,2)* model. Third, the *BP(0.5,0.5)* model is impractical for trial design because the associated post-study odds thresholds demonstrating strong equipoise imbalance require very low false positive rates and very high power, which may lead to marginal effect sizes being identified as strong clinical evidence on a statistical basis.

To demonstrate this third point, it suffices to examine what power is required by each of the models in Figure 1 to show equipoise imbalance at the 95$^{th}$ percentile of their pre-study odds distributions. The power required to achieve post study odds of 19:1 at 5% false positive rate using the *BP(1,1)* model is the unique solution to the equation $p(+|\ H_1\ )/0.05 = 19$, that is $p(+|\ H_1\ ) = 95\%$. Achieving this power would require recruitment of a larger number of study participants compared to current practice for superiority design, potentially decreasing the minimal effect sizes associated with statistically positive outcomes. Hence, a positive study outcome based on 95% power would provide strong evidence of equipoise imbalance in favour of the alternative hypothesis, but it may not necessarily be seen as improving current evidence standards based on the typical 80%-90% power range unless a minimum effect size is pre-specified.

Conversely, superiority designs carrying 95% power at 5% false positive rate demonstrate strong equipoise imbalance at the 95$^{th}$ percentile of the *BP(1,1)* model against the alternative hypothesis when the trial outcome fails to reject the null upon observing small effect sizes, which provide a more robust basis to carefully consider further clinical development decisions compared to smaller designs. Also, if the *BP(1,2)* model were to be used for equipoise calibration, the required power at 5% false positive rate would be only $p(+|\ H_1\ )$/0.05 = 3.5, that is $p(+|\ H_1\ )$ = 17.5%. Hence, calibration of trial sample size based on the *BP(1,2)* model would substantially lower clinical evidence standards compared to current practice. Finally, the *BP(0.5,0.5)* model cannot be used to calibrate trial sample size at 5% false positive rate, because the required power $p(+|\ H_1\ )$/0.05 = 161 does not admit a solution within the 0-100% range. The maximum false positive error rate associated with power below 100% under the *BP(0.5,0.5)* equipoise model is $1/161 \approx 0.62\%$. Calibration of study sample size using this false positive error rate threshold, representing and eight-fold reduction of the currently accepted 5% standard for confirmatory clinical trials, would entail large increases in sample sizes likely leading to identifying very small treatment effects as strong clinical evidence.

<table>
<tr><th colspan="2">Design operational characteristics</th><th rowspan="2">Post-study odds in favour of $H_1$</th><th rowspan="2">Pre-study <i>BP(1,1)</i> equipoise model percentile</th></tr>
<tr><th>False Positive rate</th><th>Power</th></tr>
<tr><td>10%</td><td rowspan="2">90%</td><td>9</td><td>90%</td></tr>
<tr><td>5%</td><td>18</td><td>94.74%</td></tr>
<tr><td>5%</td><td>95%</td><td>19</td><td>95%</td></tr>
<tr><td>1%</td><td>99%</td><td>99</td><td>99%</td></tr>
</table>

*Table 1: operational characteristics of the trial outcome, posterior odds in favour of the design hypothesis from observation of a positive study outcome and associated* $BP(1{,}1)$ *pre-study equipoise model percentile. The percentiles show in the right-most column are calculated as X/(1+X), for any value X of the post-study odds.*

## *2.3 Does equipoise calibration change confirmatory study design practices?*

Table 1 suggests that application of equipoise calibration may have but a small impact on sample size requirements compared to current design practices. To examine this argument in a practical setting, trial design is considered here using standard group sequential (GS) methods under strong control of the family-wise error rate (FWER) [33-35].

The design used here has one primary time-to-event (TTE) endpoint, assuming a median of 10 months in the standard of care (SOC) arm. The treatment effect size is measured by the cumulative hazard ratio (HR), following current practice. The design hypothesis $H_1$: HR = 0.7 postulates that a relative risk reduction of 30% compared to SOC will be achieved in the investigational arm, which is expected to translate into 4 months increase in median TTE under exponential distributions in both arms. The O'Brien and Fleming (OBF) approach is used to distribute the overall false positive error rate between one interim analysis (IA) at 70% information fraction and, should the trial continue beyond IA, the final analysis (FA) at 42 months follow-up. At IA and at FA respectively 36% and 52% of study participants are expected to have experienced the primary endpoint event across the two arms. This design represents assumptions for confirmatory study common in oncology, where the primary endpoint is typically progression free survival (PFS) or overall survival (OS).

Table 2 shows that, under the *BP(1,1)* equipoise model, a positive trial outcome at 90% power here provides odds of at least 19.7:1 in favour of $H_1$, which demonstrate equipoise imbalance in favour of $H_1$ at approximately the 95$^{th}$ percentile of the pre-study equipoise distributions. Also, a negative outcome here provides odds of approximately 9.5:1 in favour of $H_0$, which demonstrate equipoise imbalance in favour of $H_0$ at the 90$^{th}$ percentile of the *BP(1,1)* pre-trial equipoise distribution. Table 2 also shows that, should 21% additional participants be enrolled in the study, power would increase to 95%. This sample size increase needs to be weighed against the moderately weaker critical values, since the largest hazard ratios defining a positive study outcome would respectively increase from 0.73 to 0.75 at the interim analysis and from 0.81 to 0.82 at the final analysis. Also, in practice it may be challenging to increase the sample size by one fifth, as this increase is likely to require activation of additional sites with substantial additional costs and possibly longer timelines to study readout especially when the event rate is low, such as in the adjuvant setting or in the development of treatments against chronic diseases. Should this sample size increase be feasible, Table 2 shows that the primary trial outcome would provide equipoise imbalance at the 95$^{th}$ percentile of the *BP(1,1)* equipoise distribution regardless of whether the outcome is positive or negative as both $r_{10}(+) > 19$ and $r_{01}(-) > 19$. Establishing robust equipoise imbalance in favour of $H_0$ when a negative trial outcome is observed is valuable in practice when multiple competing compounds are concurrently developed, as a basis to consider prioritisation of further investment on compounds more likely to deliver clinically meaningful benefit.

The bottom row in Table 2 also shows that achieving 99% power is associated with marginally stronger positive evidence $r_{10}(+)$ and much stronger negative evidence $r_{01}(-)$ against equipoise. However, recruitment of 69% additional study participants entails substantial delays to study read-out and it would allow declaring a positive study outcome based on more modest hazard ratio critical values, which are unlikely to improve current clinical evidence standards.

| **Power** | **N (%)** | **HR CV**[1] | | $r_{10}(+)$ *(BP(1,1) percentile)* | | $r_{01}(-)$ *(BP(1,1) percentile)* |
|---|---|---|---|---|---|---|
| | | **IA** | **FA** | **IA** | **FA** | **FA** |
| 90% | 680 (100%) | 0.73 | 0.81 | 43.3 (97.7%) | 19.7 (95.2%) | 9.5 (90.5%) |
| 95% | 826 (121%) | 0.75 | 0.82 | 50 (98%) | 20.8 (95.4%) | 19.1 (95%) |
| 99% | 1146 (169%) | 0.79 | 0.85 | 59.7 (98.4%) | 21.7 (95.6%) | 95.4 (98.9%) |

[1] The abbreviation "CV" here stands for critical value, which is the largest hazard ratio value associated with a positive primary analysis outcome

*Table 2: power, sample size, hazard ratio critical values at 5% family-wise error rate and associated post-study odds using O'Brien and Fleming alpha spending for a group sequential design assuming a time-to-event true median of 10 months in the standard of care arm and a 30% relative risk reduction in the investigational arm.*

## 3 Equipoise calibration of clinical development plans

A recent review of late phase oncology clinical trials [36] examined the association between pre-study evidence and the primary outcome of confirmatory studies in gastro-intestinal cancers based on primary literature. Among phase 3 trials preceded by a phase 2 study designed to assess a pre-specified clinical hypothesis, the phase 3 success rate based on a positive phase 2 was roughly twice as large as for studies preceded by a statistically negative phase 2 study outcome (41.9% vs 18.9%). This observation suggests that joint design of phase 2 and phase 3 studies within one clinical development plan (CDP, [37]) is relevant to optimise the overall probability of success of clinical development in gastro-intestinal cancers. To this end, the sequential CDP design considered here comprises one randomised phase 2 trial and one phase 3 confirmatory trial. The post-study odds of the joint phase 2 and phase 3 trial hypotheses associated to the four possible CDP outcomes are respectively

$$r_{10}(+_{ph2},+_{ph3}) \coloneqq \frac{P(H_{1,ph2},H_{1,ph3} \mid +_{ph2},+_{ph3})}{P(H_{0,ph2},H_{0,ph3} \mid +_{ph2},+_{ph3})} = \frac{P(H_{1,ph2},H_{1,ph3})}{P(H_{0,ph2},H_{0,ph3})} \times \frac{p(+_{ph2} \mid H_{1,ph2})}{p(+_{ph2} \mid H_{0,ph2})} \times \frac{p(+_{ph3} \mid +_{ph2},H_{1,ph3})}{p(+_{ph3} \mid +_{ph2},H_{0,ph3})}, \quad (5)$$

$$r_{01}(+_{ph2},-_{ph3}) \coloneqq \frac{P(H_{0,ph2},H_{0,ph3} \mid +_{2},-_{ph3})}{P(H_{1,ph2},H_{1,ph3} \mid +_{ph2},-_{ph3})} = \frac{P(H_{0,ph2},H_{0,ph3})}{P(H_{0,ph2},H_{1,ph3})} \times \frac{p(+_{ph2} \mid H_{0,ph2})}{p(+_{ph2} \mid H_{1,ph2})} \times \frac{p(-_{ph3} \mid +_{ph2},H_{0,ph3})}{p(-_{ph3} \mid +_{ph2},H_{1,ph3})}, \quad (6)$$

$$r_{10}(-_{ph2},+_{ph3}) \coloneqq \frac{P(H_{1,ph2},H_{1,ph3} \mid -_{ph2},+_{ph3})}{P(H_{0,ph2},H_{0,ph3} \mid -_{ph2},+_{ph3})} = \frac{P(H_{1,ph2},H_{1,ph3})}{P(H_{0,ph2},H_{0,ph3})} \times \frac{p(-_{ph2} \mid H_{1,ph2})}{p(-_{ph2} \mid H_{0,ph2})} \times \frac{p(+_{ph3} \mid -_{ph2},H_{1,ph3})}{p(+_{ph3} \mid -_{ph2},H_{0,ph3})}, \quad (7)$$

$$r_{01}(-_{ph2},-_{ph3}) \coloneqq \frac{P(H_{0,ph2},H_{0,ph3} \mid -_{ph2},-_{ph3})}{P(H_{1,ph2},H_{1,ph3} \mid -_{ph2},-_{ph3})} = \frac{P(H_{0,ph2},H_{0,ph3})}{P(H_{0,ph2},H_{1,ph3})} \times \frac{p(-_{ph2} \mid H_{0,ph2})}{p(-_{ph2} \mid H_{1,ph2})} \times \frac{p(-_{ph3} \mid -_{ph2},H_{0,ph3})}{p(-_{ph3} \mid -_{ph2},H_{1,ph3})}. \quad (8)$$

Equations (5)-(6) represent the post-study odds in favour of the joint trial hypotheses $(H_{1,ph2},H_{1,ph3})$ or $(H_{0,ph2},H_{0,ph3})$ should the phase 2 outcome be statistically positive and the following phase 3 study outcome be respectively positive or negative. Similarly, equations (7)-(8) define the post-study odds respectively in favour of the join hypothesis $(H_{1,ph2},H_{1,ph3})$ or $(H_{0,ph2},H_{0,ph3})$ should the phase 2 outcome be statistically negative and the following phase 3 study outcome be respectively positive or negative. Cases (7)-(8) are considered because, in practice, a phase 3 trial may be initiated regardless of the phase 2 efficacy outcome achieving statistical significance, including instances when a favourable phase 2 safety profile is observed together with a phase 2 efficacy endpoint weakly associated with phase 3 design. This may be the case for phase 2 designs using binary efficacy endpoints, such as objective response rate (ORR) or differences in progression free survival probabilities (PFS) at set landmarks and phase 3 designs using overall survival (OS) [38] primary endpoint, as shown in the following example.

### *3.1 Equipoise calibration of CDP designs requires defining between-study dependence*

Equipoise calibration based on the post-study odds (5)-(8) guides the determination the CDP sample size and its distribution between phase 2 and phase 3 trials. CDP calibration ensures that positive outcomes will demonstrate strong equipoise imbalance in favour of $(H_{1,ph2}, H_{1,ph3})$ and quantifies the level of equipoise imbalance in favour of $(H_{0,ph2}, H_{0,ph3})$ associated with negative phase 3 outcomes, as a basis for further investment decisions. Joint pre-study equipoise is used here, as represented here by two independent *BP(1,1)* models. This joint equipoise model encodes minimal information about the pre-study likelihood of the study hypotheses, and it will be useful in practice only if the phase 2 and phase 3 efficacy endpoints are uncorrelated. Under this model, Table 3 shows that the post-study odds threshold 66:1 is appropriate to establishing robust equipoise imbalance about the joint phase 2 and phase 3 hypotheses from the observed CDP outcomes.

| **Joint *BP(1,1)* model percentiles** | 50% | 80% | 85% | 90% | 95% | 97.5% |
|---|---|---|---|---|---|---|
| **Odds Threshold** | 1 | 8 | 13 | 24 | 66 | 167 |

*Table 3: percentiles of the joint BP(1,1) equipoise model, for 2-trials CDP equipoise calibration. Odds thresholds were calculated as the value of the product of two independent BP(1,1) pseudo-random variates having exact Monte Carlo cumulative frequency equal to the reported percentile.*

### *3.2 Case study: equipoise calibration of an oncology 2-study CDP*

A key objective of phase 2 studies in oncology is to inform robust go/no-go decisions based on endpoints observable within relatively short timelines. To this end, phase 3 go/no-go is based here on the difference between the proportions of participants alive and at risk of progression at 9 months follow-up (PFS9) observed in the phase 2 study. Phase 2 participants are randomised 1:1 to parallel investigational and SOC arms. The phase 2 design hypothesis is $H_{1,ph2}: PFS9_{inv} = 75\%, PFS9_{soc} = 55\%$ and the primary analysis outcome is defined positive if the p-value of the chi square statistic for the observed PFS9 difference is less than 5%. The phase 3 design illustrated in Section 2 is used here, with primary endpoint based on OS. Table 4 summarises the operational characteristics and sample sizes of four CDPs, enrolling a broad range of participants. The phase 3 FWER is set at 5% for all but the most robust CDP option in Table 4, which sets the FWER to 1%. The 2-sided false positive rates of the phase 3 interim and final analyses are respectively 1.48% and 4.56% for all but the more robust CDP option listed in Table 4, where they are 0.16% and 0.94%. Of note, the total alpha spent across the two pre-planned phase 3 analyses exceeds the nominal 5% FWER due to the positive correlation between the interim and final analysis p-values.

The "Minimal" and "Upfront" CDPs in Table 4 comprise a phase 2 study sized respectively at 10% (N ~ 100) or 5% (N ~ 200) false positive rate, followed by a phase 3 design with 80% power (N=526). Table 4 shows that equipoise imbalance in favour of $(H_{1,ph2}, H_{1,ph3})$ provided by two positive trial outcomes under these two designs is strong, as $r_{10}(+_{ph2}, +_{ph3}) = \frac{0.8}{0.1} \times \frac{0.8}{0.0456} \approx 140 \gg 66$. However, evidence in favour of $(H_{0,ph2}, H_{0,ph3})$ associated to a mixed outcome is insufficient for both the "Minimal" and the "Upfront" designs, as $r_{01}(+_{ph2}, -_{ph3}) = \frac{0.1}{0.8} \times \frac{(1-0.0456)}{0.2} \approx 0.6 < 1$. This result may be surprising, as the negative outcome of the larger phase 3 trial may be expected to dominate the overall equipoise imbalance. This is not the case here, due to the operational characteristics associated with the positive outcome in phase 2 being stronger than those of the negative evidence in phase 3 so that, when these are combined, the phase 2 outcome dominates the overall result.

Specifically, the "Minimal" and "Upfront" CDP designs in Table 4 are associated with a phase 3 negative likelihood ratio of $\frac{p(-_{ph3}|H_{0,ph3})}{p(-_{ph3}|H_{1,ph3})} = \frac{1-0.0456}{0.2} = 4.77$ and with a phase 2 positive likelihood ratio respectively equal to $\frac{p(+_{ph2}|H_{0,ph2})}{p(+_{ph2}|H_{1,ph2})} = 0.1/0.8 = 0.125$ and $0.05/0.9 \approx 0.055$. Since the combined post-study odds for these two designs are $r_{01}(+_{ph2}, -_{ph3}) \leq 4.77 \times 0.125 \approx 0.6$, the overall evidence from the mixed outcome $(+_{ph2}, -_{ph3})$ numerically favours the alternative $(H_{1,ph2}, H_{1,ph3})$ despite observing a negative outcome in the larger phase 3 trial. Hence, these two CDPs do not represent viable strategies from a statistical perspective. This is not the case for the CDPs labelled "Base" in Table 4, which comprises a phase 2 design carrying 80% power at 10% false positive rate, followed by a phase 3 trial design with 90% cumulative power at 5% FWER. This strategy, which involves but a modest increase in sample size compared to the "Upfront" option, provides robust equipoise imbalance should two positive outcomes be observed, as $r_{10}(+_{ph2}, +_{ph3}) > 158 \gg 66$. The "Base" CDP design also provides weak evidence in favour of the overall null hypothesis should the phase 3 trial be negative ($r_{01}(+_{ph2}, -_{ph3}) = 1.2$) and it provides evidence in favour of the combined null above the 90th percentile of the joint equipoise model should both phase 2 and phase 3 studies be negative ($r_{01}(-_{ph2}, -_{ph3}) = 43$).

The weak level of negative evidence associated to the mixed outcome $(+_{ph2}, -_{ph3})$ for the "Base" design suggests an opportunity to explore stronger CDP strategies, represented in Table 4 by two "Robust" designs. These two options respectively raise the phase 3 power to 95% and to 99% at a lower FWER of 1%. The top "Robust" option in Table 4, carrying 80% power at 10% false positive rate in phase 2 and 95% cumulative power at 5% FWER in phase 3, is the smallest CDP ensuring that either double positive or double negative trial outcomes enable rejecting pre-CDP equipoise at the 95th percentile of the pre-study joint equipoise distribution shown in Table 3, as $r_{10}\left(+_{ph2},+_{ph3}\right) \geq 167 \gg 66$ and $r_{01}\left(-_{ph2},-_{ph3}\right) \geq 86 > 66$. The bottom row in Table 4 shows that, should the mixed CDP outcome $(+_{ph2},-_{ph3})$ be observed, equipoise imbalance in favour of the combined null exceeds the 80th *BP(1,1)* percentile in Table 3 only for the most robust design ($r_{01}\left(+_{ph2},-_{ph3}\right) = 12.4$). However, the stronger evidence provided by this robust CDP design is unlikely to outweigh the operational complexity and extension to read-out times entailed by recruitment of over 600 additional participants. Hence, Table 4 shows that CDP designs using currently accepted levels of operating characteristics are appropriate to establish equipoise imbalance in favour of $(H_{1,ph2}, H_{1,ph3})$ based on positive phase 2 and phase 3 outcomes, and that more robust CDPs are needed for a negative phase 3 study to overcome a positive phase 2 outcome.

| **CDP** | **Total N**<br>**(Ph3 N)** | $p(+_{ph2} \mid H_{0,ph2})$<br>$p(+_{ph3} \mid H_{0,ph3})$ | $p(+_{ph2} \mid H_{1,ph2})$<br>$p(+_{ph3} \mid H_{1,ph3})$ | **Ph2 ORR difference CV** | **Ph3 HR CV** | | $r_{10}(+_{ph2},+_{ph3})$ | | $r_{01}(+_{ph2},-_{ph3})$ | $r_{10}(-_{ph2},+_{ph3})$ | | $r_{01}(-_{ph2},-_{ph3})$ |
|---|---|---|---|---|---|---|---|---|---|---|---|---|
| | | | | | IA | FA | IA | FA | FA | IA | FA | FA |
| Minimal | 626<br>(526) | 10%<br>5% | 80%<br>80% | 18% | 0.7 | 0.78 | 276 | 140 | 0.60 | 7.7 | 3.9 | 21 |
| Upfront | 718<br>(526) | 5%<br>5% | 90%<br>80% | 15.6% | | | 621 | 316 | 0.27 | 3.6 | 1.9 | 45 |
| Base | 780<br>(680) | 10%<br>5% | 80%<br>90% | 18% | 0.73 | 0.81 | 346 | 158 | 1.2 | 9.6 | 4.4 | 43 |
| Robust | 926<br>(826) | | 80%<br>95% | | 0.75 | 0.82 | 400 | 167 | 2.4 | 11 | 4.6 | 86 |
| | 1584<br>(1484) | 10%<br>1% | 80%<br>99% | | 0.76 | 0.83 | 4175 | 843 | 12.4 | 116 | 23 | 446 |

*Table 4: characteristics of clinical development plan designs comprising one phase 2 trial and one phase 3 trial. The designs labelled "Minimal" and "Upfront" deliver strong equipoise imbalance if both trials are positive, but insufficient evidence against the join null if the phase 3 trial fails to confirm a positive phase 2 outcome. The "Base" case delivers numerical*

*evidence against the joint null in the mixed outcome case* $(+_{ph2}, -_{ph3})$. *The two "Robust" options further increase the levels of equipoise imbalance by recruiting respectively 19% and 100% additional study participants compared to the "Base" case. The top robust design is the smallest CDP delivering strong equipoise imbalance when double positive or double negative outcomes are observed in the phase 2 and phase 3 studies.*

## Discussion

The models presented here characterize the level of uncertainty about competing clinical hypotheses at population level prior to observing the outcome of a clinical trial designed to reduce that uncertainty. The importance of a clear representation of pre-study uncertainty supporting the clinical interpretation of the strength of evidence provided by the study outcome was highlighted by the widely different power, and hence sample size, required by different models. Statistically positive outcomes under the *BP(0.5,0.5)* model were shown to be unlikely to be associated with relevant treatment effect magnitudes. The *BP(1,2)* population average model showed that distributions assigning different pre-study weight to the null and the alternative hypotheses can encode average equipoise in aggregate. However, using this distribution may introduce bias in the study design, and hence it would need stronger justifications. The *BP(1,1)* model shows that the level of equipoise imbalance provided by positive trial outcomes at traditional levels of operating characteristics for confirmatory clinical trials is greater than 90%. Further increasing power is identified as a potential avenue for trial outcomes to convey even more robust evidence of equipoise imbalance when a study outcome is to inform subsequent development plans. For instance, investigational agents may be clinically developed first as monotherapy and then in combination. Here equipoise calibration provides a formal method to determining the operational characteristics of the monotherapy phase 3 trial to ensuring that a negative outcome provides sufficient evidence to discontinue further development in absence of synergistic interactions with combination partners. The study of the potential impact of applying equipoise calibration in these specific circumstances is beyond the scope of this paper.

Although the examples examined in this work pertain mostly to designs used in oncology clinical development, equipoise calibration is equally applicable to designs and endpoints relevant to clinical development in areas other than oncology, which were not examined in this paper. A prominent example is provided by clinical development plans including phase 2 trials using early efficacy endpoints, which may be based on baseline of pharmacodynamic biomarkers positively correlated to the phase 3 efficacy endpoints, thus demonstrating potential for patient benefit from longer term exposure to investigational compounds.

The models explored here may be further developed in several directions, including more refined definitions clinical equipoise using parameters of the pre-study odds distribution beyond the median or the mean, or if pre-existing evidence were to be clearly quantifiable and relevant to the design of confirmatory trials. One such instance pertains to the design of clinical development plans where a phase 2 endpoint is correlated with the primary phase 3 endpoint, due to a shared mechanism of action in similar patient populations. In these situations, it may be appropriate to include evidence provided by the phase 2 outcome in the design and analysis of the phase 3 trial, to effectively accelerate clinical development. Here equipoise calibration provides a framework for calculating the sample size needed to achieve an overall level of strong equipoise imbalance in phase 3 from pre-trial odds other than 1:1. The study of specific circumstances where these designs may be successful in practice and the regulatory requirements for demonstrating imperfect pre-study equipoise for a primary phase 3 endpoint lie beyond the scope of this paper and provide further topics for further research.

**Declarations**

This work was completed while the author was a full-time employee of AstraZeneca PLC, which consented to its publication. No dedicated funding was needed.

No external data were used beyond the publications reported in the References Section below.

The Author declares no competing interests. The Author wishes to acknowledge the constructive comments received from Editors and Referees on a previous version of this work.

No LLM support was used during the development of this work.